**Impact of Tandem Repeats on the Scaling of Nucleotide Sequences**

Radhakrishnan Nagarajan*

*Center on Aging, University of Arkansas for Medical Sciences*

Meenakshi Upreti

*Department of Biochemistry and Molecular Biology, University of Arkansas for Medical Sciences*

*To whom correspondence should be addressed*

Radhakrishnan Nagarajan

Center on Aging, University of Arkansas for Medical Sciences

629 Jack Stephens Drive, Room: 3105

Little Rock, Arkansas 72205

Phone: (501) 526 7461

Email: nagarajanradhakrish@uams.edu

**ABSTRACT**

Techniques such as detrended fluctuation analysis (DFA) and its extensions have been widely used to determine the nature of scaling in nucleotide sequences. In this brief communication we show that tandem repeats which are ubiquitous in nucleotide sequences can prevent reliable estimation of possible long-range correlations. Therefore, it is important to investigate the presence of tandem repeats prior to scaling exponent estimation.

**1. Introduction**

Understanding patterns in eukaryotic DNA sequences is an area of active research. This is more so with the rapid completion of eukaryotic genomes. DNA sequences are composed of four nucleotides (A, G, C and T), with (A, G) representing the *purines* and (C, T) the *pyrimidines*. Repetitive nucleotide patterns form a prominent part of eukaryotic genomes and manifest themselves as *tandem repeats*. Such repeats are usually approximate repeats adjacent to each other and include microsatellites, minisatellites, CpG islands, and telomeric repeats [Dogett et al., 1992; Toth et al., 2000]. Classical techniques such a Fourier analysis have been used to identify short-term correlated patterns in DNA sequences [Silverman & Linsker, 1986; Tavare & Giddings, 1989; Coward, 1997]. Such correlations are of finite memory or *Markovian* in nature, and can be broadly classified into periodic and quasi-periodic repeats. However, recent studies have provided compelling evidence of *non-Markovian* characteristics in the form of *long-range correlations* (LRC) in DNA sequences [Peng et al., 1992; Li and Kaneko, 1992]. Such correlations persist over large nucleotide distances, also known as time scales. This often results in power-law spectral signatures of the form $S(f) \sim f^{-b}$, where frequency is represented as the reciprocal of basepairs (bp), i.e. (1/bp). Several algorithms have been proposed in the past to determine the scaling exponent of a given sequence [Feder, 1988; Li & Kaneko, 1992; Peng et al., 1992]. *Detrended fluctuation analysis* DFA and its extensions multifractal DFA (MF-DFA) [Peng et al., 1992; Kantelhardt et al., 2002] have been successfully used in the past to estimate the scaling exponents in sequences obtained from diverse settings (Hu et al., 2001 and references there in). DNA sequences are composed mainly of coding regions (*exons*) which code for specific proteins and non-coding regions (*introns*) which are interspersed between exons. Earlier studies provided overwhelming evidence of LRCs in the introns ($\alpha > 0.5$) and absence of correlations in exons ($\alpha = 0.5$) [Li & Kaneko, 1992; Peng et al., 1992]. Subsequently, several independent studies have used DFA to arrive at similar results.

 In the present study, we show that estimation of the scaling exponent can be significantly affected by the presence of tandem repeats which are ubiquitous in DNA sequences. While scaling in introns have been associated with the presence of long-range correlations ($\alpha > 0.5$) drawing such conclusion can be non-trivial in the presence of tandem repeats. This is attributed to characteristic distortion introduced by the tandem repeats in the log-log plot of the fluctuation function versus time scale, rendering the estimation of the scaling exponent unreliable. Therefore, it is important to examine a given nucleotide sequence for tandem repeats prior to scaling exponent estimation. The present study also raises new concerns on interpreting results obtained on genes and chromosomes, which are riddled with such repeats.

Recent reports have investigated the impact of stationary uncorrelated trends superimposed on power-law noise [Hu et al., 2001; Nagarajan & Kavasseri, 2005; Nagarajan & Kavasseri, 2005]. However, there are subtle differences between stationary uncorrelated trends and tandem repeats. Tandem repeats occur as patches hence non-stationary in the sequence space. Such repeats can also be an inherent part of the dynamics, hence may not fall under the class of uncorrelated trends. In a recent study, GC rich tandem repeats have been found to exhibit oscillations characteristic of deterministic nonlinear dynamical systems [Nicolay et al., 2004]. Therefore, extending earlier results [Hu et al., 2001; Nagarajan & Kavasseri, 2005; Nagarajan & Kavasseri, 2005] on the impact of uncorrelated trends on power-law noise to tandem repeats is not immediate.

## 2. Methods

The given nucleotide sequence of alphabet size four (A, G, C, T) is converted into an indicator sequence with alphabet size two by the mapping (A, G) $\rightarrow$ -1 and (C, T) $\rightarrow$ 1 [Peng et al., 1992]. A brief description of the DFA (MF-DFA) procedure is enclosed below for completeness. A detailed treatment can be found elsewhere [Peng et al., 1992; Kantelhardt et al., 2002].

Description of DFA and MF-DFA:

a.  Determine the integrated series from the given indicator sequence $\{x_k\}, k = 1...n$ as

$$y_i = \sum_{k=1}^{i} (x_k - \boldsymbol{m}), \ i = 1...n \text{ where } \boldsymbol{m} \text{ represents the average value of } \{x_k\}, k = 1...n$$

b.  The data is divided into $n_s$ non-overlapping bins of size $s$, where $n_s = \left\lfloor \dfrac{n}{s} \right\rfloor$ ; $\lfloor \ \rfloor$

represents the greatest integer not larger than $\dfrac{n}{s}$. The local trend $y_{\boldsymbol{u}}(i)$ is calculated is

calculated in each of the bins ($\boldsymbol{u} = 1...n_s$) by a polynomial regression and the variance is

determined, given by

$$F^2(\boldsymbol{u}, s) = \frac{1}{s} \sum_{i=1}^{s} \left\{ y[(\boldsymbol{u}-1)s + i] - y_{\boldsymbol{u}}(i) \right\}^2$$

Recent studies have suggested higher order polynomial regression in order to minimize the effect of local trends [Ashkenazy et al., 2001; Kantelhardt et al., 2001].

c.  This procedure is repeated from either end of the data in order to accommodate all the

samples. Thus the effective length of the data is $2n_s$.

d.  The fluctuation $F^2(\boldsymbol{u}, s)$ is averaged over to obtain the $q^{th}$ order fluctuation function

given by

$$F_q(s) = \left\{ \frac{1}{2n_s} \sum_{\boldsymbol{u}=1}^{2n_s} F^2(\boldsymbol{u}, s)^{q/2} \right\}^{1/q}$$

where $q$ can take any real value $(q \neq 0)$. The scaling behavior is determined by analyzing the log-log plots of $F_q(s)$ versus $s$.

The DFA code for estimating the scaling exponent is publicly available and can be obtained from (http://www.physionet.org/physiotools/dfa/).

The scaling exponent(s) of $\{x_k\}, k = 1 \ldots n$ is determined from the log-log plot of fluctuation function $F_q(s)$ versus time scale ($s$) with varying moments ($q$, $q \in R$ and $q \neq 0$). Monofractal sequences exhibit a single exponent and their fluctuation function varies *linearly* with time scale in the log scale. Therefore, determining the variation of $F_q(s)$ versus ($s$) for a single moment (q = 2) is sufficient to capture their scaling behavior. This has to be contrasted with multifractals whose fluctuation function varies as a function of moments $q$. It should be noted that investigating the qualitative dynamics of the fluctuation function versus time scale with varying moments can be useful in discerning monofractal from multifractal dynamics in the presence of trends [Nagarajan and Kavasseri, 2005].

To establish the effect of tandem repeats on scaling exponent estimate, we compare the scaling behavior of introns *with tandem repeats* such as (CpG islands) [Gardiner-Garden & Frommen, 1987; Aissani & Bernardi, 1991; Larsen et al., 1992] to those *without tandem repeats*. CpG islands are GC rich regions which are resistant to methylation, hence directly associated with the activity of genes. Such regions are ubiquitous in house keeping and tissue specific genes, hence their choice is valid within the present context. We consider introns from (Human blood coagulation factor VII (HUMCFVII, Genbank ID: J02933, 12850 basepairs) [O'Hara et al., 1987] and Homo Sapiens limb-girdle muscular dystrophy type 2b gene (LGMD2B, Genbank ID: AJ007973, 36133 basepairs) [Bashir et al., 1998]. The choice of HUMCFVII is encouraged by Li and Kaneko's seminal report on the existence of partial power-law decay and LRCs in this gene. HUMCFVII has nine exons and eight introns located at (522:585; 1654:1719; 4294:4454; 6383:6407; 6478:6591; 8307:8447; 9419:9528; 10124:10247; 11064:11659) and (586:1653;

1720:4293; 4455:6382; 6408:6477; 6592:8306; 8448:9418; 9529:10123; 10248:11063) respectively. Investigating HUMCFVII for possible GC rich domains (CpG plot, EMBOSS, http://www.ebi.ac.uk/emboss/cpgplot/ with parameters Observed/Expected ratio = 0.6; length > 200; Percent C + Percent G = 50%) revealed three distinct regions: (2075:2677), (2996:4726), (11050:11313) with lengths 603, 1731 and 264 respectively. Two of the GC rich domains (2075:2677 and 2996:4726) had a significant overlap with the intron located at (1720:4293). In the subsequent discussion we shall refer to this GC rich intron as (Intron$_{REP}$, N = 2574 nucleotides). LGMD2B gene consists of three exons and three introns located at (19078:19176; 31270:31398; 33986:34141; 35668:35759) and (19177:31269; 31399:33985; 34142:35667) respectively. Investigating the intron (19177:31269) using CpG plot with the same parameters as above failed to reveal any GC rich domains. Therefore, we use this intron as the reference sequence. In order reject the claim that the observed distortion (crossover) in scaling is an outcome of finite sample size effects, we chose the region (19177+1720:19177+4293) which is of the same length as Intron$_{REP}$ (N = 2574). Since this intron does not contain any GC rich regions we shall refer to it as (Intron$_{NOREP}$) in the subsequent discussion.

## 3. Results

Spectral signatures of Intron$_{NOREP}$ and Intron$_{REP}$ exhibited characteristic low-frequency power-law decay, Fig. 1, conforming to earlier observations of LRC in introns [Li & Kaneko., 1992; Peng et al., 1992]. In addition, Intron$_{REP}$ exhibited prominent peaks which were absent in Intron$_{NOREP}$, Fig. 1. These confirm our earlier observation of patchy GC rich domains in Intron$_{REP}$ unlike Intron$_{NOREP}$. Investigating the log-log plot of the fluctuation function $F_2(s)$ versus sequence length (s) revealed characteristic crossover for Intron$_{REP}$ indicating possible existence of more than one scaling exponent, Fig. 2a. The nonlinear nature of the log-log plot also prevented reliable estimation of the scaling exponent. This has to be contrasted with that of Intron$_{NOREP}$ whose log-log plot was linear with a scaling exponent ($\alpha \sim 0.6$). In order to reject the claim that the crossover in the case of Intron$_{REP}$ is due to local polynomial trends in the integrated series, $\{y_k\}, k = 1...n$ , Sec. 2, we invest investigated the log-log plot of F$_2$(s) versus (s) of Intron$_{NOREP}$ and Intron$_{REP}$ with various orders of polynomial detrending (d = 1, 2, 3 and 4), Fig. 2. The log-log plots with (d = 1, 2, 3 and 4) failed to reveal any characteristic change in the scaling behavior indicating that the observed crossover is not an outcome of local polynomial trends in the integrated series.

The results obtained using DFA on Intron$_{NOREP}$ and Intron$_{REP}$ were also confirmed with more classical tools such as rescaled-range (R/S) analysis [Feder, 1988], Fig. 3. While the scaling of Intron$_{NOREP}$ was linear with exponent ($\alpha \sim 0.6$) those of Intron$_{REP}$ was nonlinear with a marked crossover, Fig. 2.

## 4. Discussion

In this brief note, we showed that tandem repeats can have a significant impact on reliable estimation of possible long-range correlations in DNA sequences. Such repeats can introduce crossovers and nonlinearity in scaling of nucleotide sequences even at the level of introns. Several reports have been published in the past on the scaling behavior of introns, exons, genes and chromosomes. However, such sequences are riddled with various types of tandem repeats. The present study encourages investigation of tandem repeats as an important preliminary step prior to scaling exponent estimation.

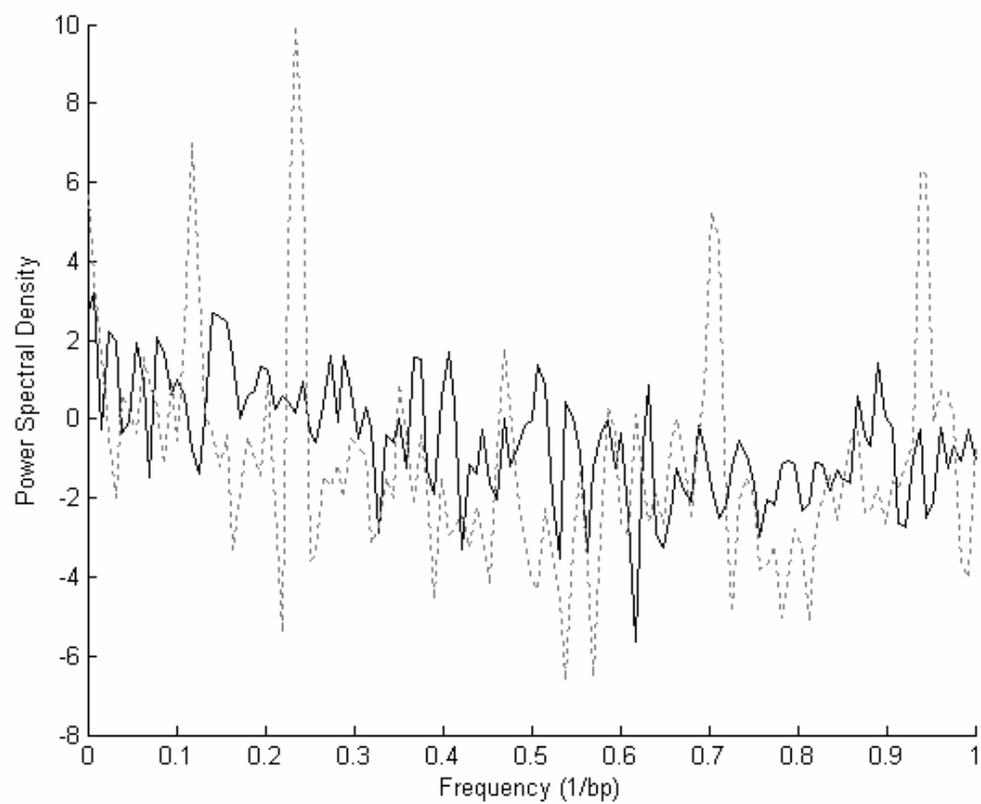

**Figure 1** Power spectral density of Intron$_{REP}$ (dotted line) and Intron$_{NOREP}$ (solid line).

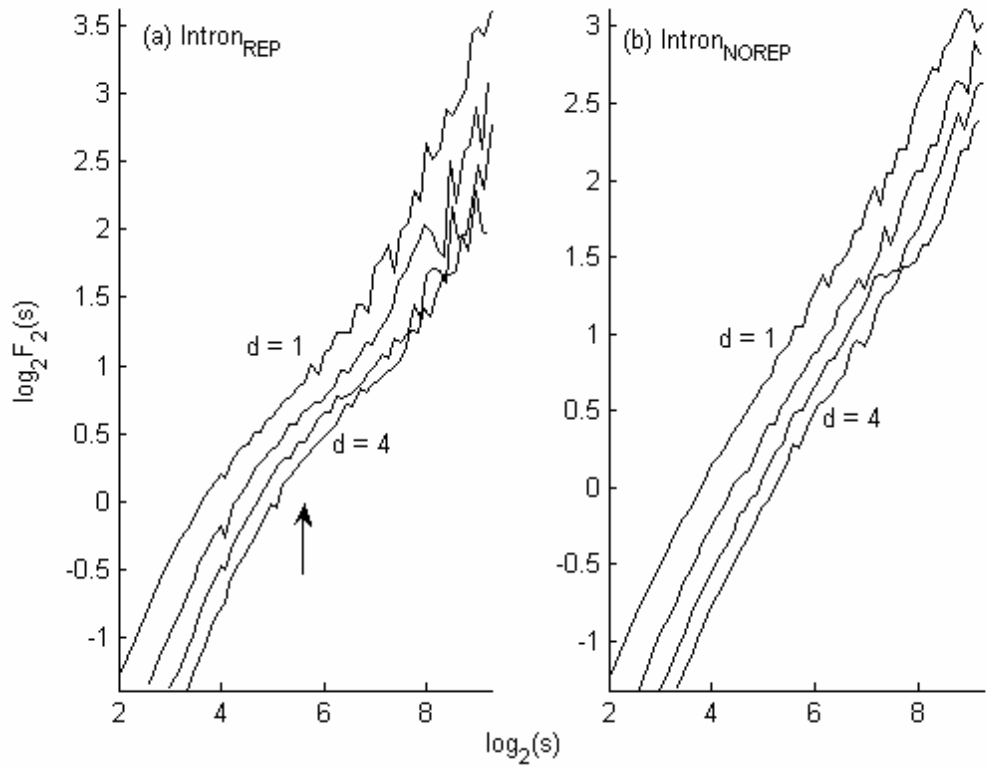

**Figure 2** Log-log plot ($\log_2$ scale) of the fluctuation function $F_2(s)$ versus sequence length ($s$) of

$\text{Intron}_{\text{REP}}$ (a) and $\text{Intron}_{\text{NOREP}}$ (b) with various orders of polynomial detrending ($d = 1, 2, 3$ and $4$, from top to bottom, in that order). The arrow in (a) indicates crossover in the scaling behavior.

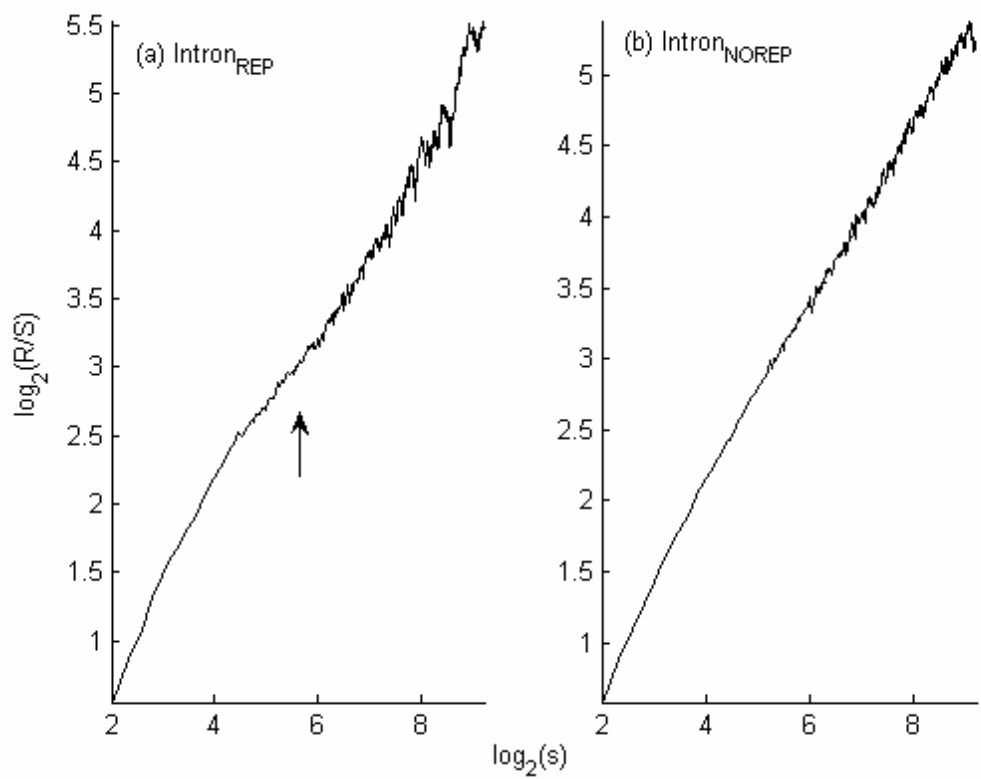

**Figure 3** Log-log plots ($\log_2$) of (R/S) versus sequence length (s) of $Intron_{NOREP}$ (a) and $Intron_{REP}$ (b) obtained using rescaled-range analysis. The arrow in (a) indicates crossover in the scaling behavior.

**Reference**

Aissani, B., & Bernardi, G. [1991] "CpG islands: features and distribution in the genomes of vertebrates," Gene 106(2), 173-183.

Ashkenazy, Y., Ivanov, P.Ch., Havlin, S., Peng, C-K., Goldberger, A.L. & Stanley, H.E. [2001] "Magnitude and sign correlations in heartbeat fluctuations," Phy. Rev. Lett. 86(9), 1900-1903.

Bashir,R., Britton,S., Strachan,T., Keers,S., Vafiadaki,E., Lako,M., Richard,I., Marchand,S., Bourg,N., Argov,Z., Sadeh,M., Mahjneh,I., Marconi,G., Passos-Bueno,M.R., de S Moreira,E., Zatz,M., Beckmann,J. & Bushby,K. [1998] "A gene related to Caenorhabditis elegans spermatogenesis factor fer-1 is mutated in limb-girdle muscular dystrophy type 2B," Nature Genetics 20 (1), 37-42.

Bassingthwaite, J.B., Liebovitch, L.S. & West, B.J. [1994] Fractal physiology New York, NY: Oxford University Press.

Coward, E. [1997] "Equivalence of two Fourier methods for biological sequences," J. Math. Biol. 36, 64-70.

Doggett, N.A., Raymond, R.L., Hildebrand, C.E. & Moyzis, R.K. [1992] "The Mapping of Chromosome," Los Alamos Science 20:182-198.

Gardiner-Garden, M. & Frommer, M. [1987] "CpG islands in vertebrate genomes," J Mol Biol. 196(2), 261-82.

Hu, K., Ivanov, P.C.h., Chen, Z., Carpena, P. & Stanley H. E. [2001] "Effect of trends on detrended fluctuation analysis," Phys. Rev. E. 64, 011114 (19 pages).

Kantelhardt, J.W., Koscielny-Bunde, E., Rego, H.A., Havlin, S. and Bunde, A. [2001] "Detecting long-range correlations with detrended fluctuation analysis," Physica A 295, 441.

Kantelhardt, J.W., Zschiegner, S.A., Koscielny-Bunde, E., Havlin, S., Bunde A. and Stanley, H. E. [2002] "Multifractal detrended fluctuation analysis of nonstationary time series," Physica A 316, 87.

Larsen, F., Gundersen, G., Lopez L. & Prydz, H. [1992] "CpG island as Gene Markers in the Human Genome," Genomics," 13, 1095-1107.

Li, W. & Kaneko, K. [1992] "Long range correlations and partial $1/f^a$ spectrum in noncoding DNA sequence," Europhys. Lett. 17(7), 655-660.

Nagarajan, R. & Kavasseri, R.G. [2005] "Minimizing the effect of sinusoidal trends on deterended fluctuation analysis," Int. J. Bifurcation & Chaos 15(2), 1767-1773.

Nagarajan, R. & Kavasseri, R.G. [2005] "Minimizing the effect of trends on deterended fluctuation analysis of long-range correlated noise," Physica A 354, 182-198.

Nicolay, S., Argoul, F., Touchon, M., d'Aubenton-Carafa, Y., Thermes, C. & Arneodo, A. [2004] "Low frequency rhythms in human DNA sequences: a key to the organization of gene location and orientation?," Phys. Rev. Lett. 93(10), 108101.

O'Hara,P.J., Grant,F.J., Haldeman,B.A., Gray,C.L., Insley,M.Y., Hagen,F.S. & Murray, M.J. [1987] "Nucleotide sequence of the gene coding for human factor VII, a vitamin K-dependent protein participating in blood coagulation," Proc. Natl. Acad. Sci. U.S.A. 84 (15), 5158-5162.

Peng, C-K., Buldyrev, S.V., Goldberger, A.L., Havlin, S., Sciortino, F., Simon, M. & Stanley, H.E. [1992] "Long range correlations in Nucleotide Sequences," Nature. 356, 168-170.

Silverman, B.D. & Linsker, R. [1986] "A measure of DNA periodicity," J. Theor. Biol. 118, 295-300.

Tavare, S. & Giddings, B.W. [1989] "Some statistical aspects of the primary structure of nucleotide sequences. Mathematical Methods for DNA Sequences," Boca Raton. FL CRC Press.

Toth, G., Gaspari, Z. & Jurka, J. [2000] "Microsatellites in different eukaryotic genomes: survey and analysis," Genome Res. 10(7), 967-81.